\newcommand*{\QUANTUM}{}%
\ifdefined\QUANTUM
\documentclass[a4paper,onecolumn,11pt,accepted=2020-12-04]{quantumarticle}
\pdfoutput=1
\else
\documentclass[11pt]{article}
\fi

\usepackage{geometry}
\usepackage{amsmath}
\usepackage{amsfonts}
\usepackage{amssymb}
\usepackage{amsthm}
\usepackage{braket}
\usepackage[qm]{qcircuit}
\usepackage{enumitem}
\usepackage{color}
\usepackage{multirow}
\usepackage[sort,square,numbers]{natbib}
\usepackage{cellspace} 
\usepackage{makecell} 
\setcellgapes{3pt}

\usepackage{hyperref}

\newtheorem{thm}{\protect\theoremname}
\theoremstyle{plain}
\newtheorem{lem}[thm]{\protect\lemmaname}
\theoremstyle{plain}
\newtheorem{rem}[thm]{\protect\remarkname}
\theoremstyle{plain}
\newtheorem*{lem*}{\protect\lemmaname}
\theoremstyle{plain}

\theoremstyle{plain}
\newtheorem{cor}[thm]{\protect\corollaryname}

\usepackage[USenglish]{babel}
  \providecommand{\corollaryname}{Corollary}
  \providecommand{\lemmaname}{Lemma}
  \providecommand{\propositionname}{Proposition}
  \providecommand{\remarkname}{Remark}
\providecommand{\theoremname}{Theorem}
\usepackage[normalem]{ulem}

\usepackage{algorithm}
\usepackage{algorithmic}

\newcommand{\SWITCH}[1]{\STATE \textbf{switch} (#1)}
\newcommand{\ENDSWITCH}{\STATE \textbf{end switch}}
\newcommand{\CASE}[1]{\STATE \textbf{case} #1\textbf{:} \begin{ALC@g}}
\newcommand{\ENDCASE}{\end{ALC@g}}
\newcommand{\CASELINE}[1]{\STATE \quad \textbf{case} #1\textbf{:} }
\newcommand{\DEFAULT}{\STATE \textbf{default:} \begin{ALC@g}}
\newcommand{\ENDDEFAULT}{\end{ALC@g}}
\newcommand{\DEFAULTLINE}[1]{\STATE \textbf{default:} }

\title{Near-optimal ground state preparation}
\author{Lin Lin}
        \affiliation{Department of Mathematics, University of California, Berkeley,  CA 94720, USA}
        \affiliation{Computational Research Division, Lawrence Berkeley National Laboratory, Berkeley, CA 94720, USA}
        \orcid{0000-0001-6860-9566}

\author{Yu Tong}
        \affiliation{Department of Mathematics, University of California, Berkeley,  CA 94720, USA}
        \orcid{0000-0002-7555-9373}

\newcommand{\apriori}{\textit{a priori }}

\newcommand{\sign}{\mathrm{sign}}
\newcommand{\poly}{\mathrm{poly}}
\newcommand{\REF}{\mathrm{REF}}
\newcommand{\PROJ}{\mathrm{PROJ}}

\begin{document}

\global\long\def\ve{\varepsilon}
\global\long\def\R{\mathbb{R}}
\global\long\def\Rn{{\mathbb{R}^{n}}}
\global\long\def\Rd{{\mathbb{R}^{d}}}
\global\long\def\E{\mathbb{E}}
\global\long\def\P{\mathbb{P}}
\global\long\def\bx{\mathbf{x}}
\global\long\def\vp{\varphi}
\global\long\def\ra{\rightarrow}
\global\long\def\smooth{C^{\infty}}
\global\long\def\symm{\mathcal{S}^n}
\global\long\def\psd{\mathcal{S}^n_{+}}
\global\long\def\pd{\mathcal{S}^n_{++}}
\global\long\def\dom{\mathrm{dom}\,}
\global\long\def\intdom{\mathrm{int}\,\mathrm{dom}\,}
\global\long\def\Tr{\mathrm{Tr}}

\newcommand{\LL}[1]{\textcolor{blue}{[LL:#1]}}
\newcommand{\YT}[1]{\textcolor{cyan}{[YT:#1]}}

\newcommand{\bvec}[1]{\mathbf{#1}}
\renewcommand{\Re}{\operatorname{Re}}
\renewcommand{\Im}{\operatorname{Im}}
\newcommand{\textred}[1]{\textcolor{red}{#1}}

\newcommand{\mc}[1]{\mathcal{#1}}
\newcommand{\mf}[1]{\mathfrak{#1}}
\newcommand{\mcV}{\mathcal{V}}
\newcommand{\Vin}{V_{\mathrm{in}}}
\newcommand{\Vstar}{V^{\ast}}
\newcommand{\Jstar}{J_{\ast}}
\newcommand{\tJstar}{\wt{J}_{\ast}}
\newcommand{\Vout}{V_{\mathrm{out}}}
\newcommand{\RPA}{\mathrm{RPA}}
\newcommand{\xc}{\mathrm{xc}}
\newcommand{\vr}{\bvec{r}}
\newcommand{\vF}{\bvec{F}}
\newcommand{\vg}{\bvec{g}}
\newcommand{\vR}{\bvec{R}}
\newcommand{\vq}{\bvec{q}}
\newcommand{\vx}{\bvec{x}}
\newcommand{\ud}{\,\mathrm{d}}
\newcommand{\ext}{\mathrm{ext}}
\newcommand{\KS}{\mathrm{KS}}
\newcommand{\Exc}{E_{\mathrm{xc}}}
\newcommand{\Vxc}{\hat{V}_{\mathrm{xc}}}
\newcommand{\Vion}{\hat{V}_{\mathrm{ion}}}
\newcommand{\abs}[1]{\lvert#1\rvert}
\newcommand{\norm}[1]{\lVert#1\rVert}
\newcommand{\average}[1]{\left\langle#1\right\rangle}
\newcommand{\wt}[1]{\widetilde{#1}}
\newcommand{\hxc}{\mathrm{hxc}}

\newcommand{\etc}{\textit{etc.}~}
\newcommand{\etal}{et al.~}  
\newcommand{\ie}{i.e.~}
\newcommand{\eg}{\textit{e.g.}~}
\newcommand{\Or}{\mathcal{O}}
\newcommand{\mcF}{\mathcal{F}}
\newcommand{\lmin}{\lambda_{\min}}
\newcommand{\lmax}{\lambda_{\max}}
\newcommand{\Ran}{\text{Ran}}
\newcommand{\I}{\mathrm{i}} 
\newcommand{\EE}{\mathbb{E}}
\newcommand{\NN}{\mathbb{N}}
\newcommand{\RR}{\mathbb{R}}
\newcommand{\CC}{\mathbb{C}}
\newcommand{\ZZ}{\mathbb{Z}}
\newcommand{\Hper}{H^1_\#(\Omega)}
\newcommand{\jmp}[1]{\jl#1\jr}
\newcommand{\al}{\{\hspace{-3.5pt}\{}
\newcommand{\avg}[1]{\al#1\ar}
\newcommand{\jl}{[\![}
\newcommand{\jr}{]\!]}
\newcommand{\VN}{\mathbb V_N}
\newcommand{\angstrom}{\mbox{\normalfont\AA}~}

\maketitle

\begin{abstract}
Preparing the ground state of a given Hamiltonian and estimating its ground energy are important but computationally hard tasks. However, given some additional information, these problems can be solved efficiently on a quantum computer. We assume that an initial state with non-trivial overlap with the ground state can be efficiently prepared, and the spectral gap between the ground energy and the first excited energy is bounded from below. With these assumptions we design an algorithm that prepares the ground state when an upper bound of the ground energy is known, whose runtime has a logarithmic dependence on the inverse error. 
When such an upper bound is not known, we propose a hybrid quantum-classical algorithm to estimate the ground energy, where the dependence of the number of queries to the initial state on the desired precision is exponentially improved compared to the current state-of-the-art algorithm proposed in [Ge et al. 2019]. These two algorithms can then be combined to prepare a ground state without knowing an upper bound of the ground energy. We also prove that our algorithms reach the complexity lower bounds by applying it to the unstructured search problem and the quantum approximate counting problem. 
\end{abstract}

\section{Introduction}

Estimating ground energy and obtaining information on the ground state of a given quantum Hamiltonian are of immense importance in condensed matter physics, quantum chemistry, and quantum information. 
Classical methods suffer from the exponential growth of the size of Hilbert space, and therefore quantum computers are expected to be used to overcome this difficulty. However even for quantum computer, estimating the ground energy is a hard problem: deciding whether the smallest eigenvalue of a generic local Hamiltonian is greater than $b$ or smaller than $a$ for some $a<b$ is QMA-complete \cite{KitaevShenVyalyi2002, KempeKitaevRegev2006, OliveiraTerhal2005, AharonovGottesmanEtAl2009}. 

Therefore to make the problem efficiently solvable we need more assumptions. We denote the Hamiltonian we are dealing with by $H$, and consider its spectral decomposition $H=\sum_k\lambda_k\ket{\psi_k}\bra{\psi_k}$ where $\lambda_{k}\leq\lambda_{k+1}$. The key assumption is that we have an initial state $\ket{\phi_0}$ which can be efficiently prepared by an oracle $U_I$, and has some overlap with the ground state $\ket{\psi_0}$ lower bounded by $\gamma$. This is a reasonable assumption in many practical scenarios. For instance, even for strongly-correlated molecules in quantum chemistry, there is often a considerable overlap between the true ground state and the Hartree-Fock state. 
The latter can be trivially prepared in the molecular orbital basis, and efficiently prepared in other basis \cite{KivlichanMcCleanEtAl2018}.
For the moment we also assume the spectral gap is bounded from below: $\lambda_1-\lambda_0\geq\Delta$.

With these assumptions we can already use phase estimation coupled with amplitude amplification \cite{BrassardHoyerMoscaEtAl2002} to prepare the ground state, if we further know the ground energy to high precision.  To our knowledge, the most comprehensive work on ground state preparation and ground state energy estimation was done by Ge \etal \cite{GeTuraCirac2019}, which provided detailed complexity estimates for well-known methods such as phase estimation, and proposed new methods to be discussed below. As analyzed in \cite[Appendix A]{GeTuraCirac2019}, in order to prepare the ground state to fidelity\footnote{In this work, the fidelity between states $\ket{x},\ket{y}$ is defined to be $\abs{\braket{x|y}}$.} $1-\epsilon$, the runtime of the controlled-time-evolution of the Hamiltonian is $\wt{\Or}(1/(\gamma^2\Delta\epsilon))$ \footnote{In this work the notation $\wt{\Or}(f)$ means $\Or(f\poly\log(f))$ unless otherwise stated.}, and the number of queries to  $U_I$ is $\wt{\Or}(1/\gamma)$, assuming the spectral norm of $H$ is bounded by a constant.  
This is however far from optimal. Poulin and Wocjan \cite{PoulinWocjan2009} proposed a method that, by executing the inverse of phase estimation to filter out the unwanted components in the initial state, can prepare a state whose energy is in a certain given range. A different choice of parameters yields a way to prepare the ground state to fidelity $1-\epsilon$ by running the controlled-time-evolution of the Hamiltonian with $\wt{\Or}(1/(\gamma\Delta)\log(1/\epsilon))$ runtime, and using  $\wt{\Or}(1/\gamma)$ queries to $U_I$ \cite[Appendix C]{GeTuraCirac2019}.

A key difference between ground state preparation and Hamiltonian simulation, where significant progress has been made in recent years \cite{Lloyd1996universal,BerryChildsKothari2015,BerryChildsCleveEtAl2015,LowChuang2017,LowWiebe2018,LowChuang2019,ChildsSuEtAl2019}, is its non-unitary nature. The recent development of linear combination of unitaries (LCU) method \cite{BerryChildsKothari2015, ChildsKothariSomma2017} provided a versatile tool to apply non-unitary operators.
Using LCU, Ge \etal proposed a new method to filter the initial state by applying a linear combination of time-evolutions of different time length \cite{GeTuraCirac2019}, which achieves the same complexity, up to logarithmic factors, as the modified version of Poulin and Wocjan's method discussed above.

All of the above methods prepare the ground state assuming the ground energy is known to high precision. When the ground energy is unknown, Ge \etal proposed a method to estimate the ground energy using a search method called minimum label finding \cite{GeTuraCirac2019}. This method can estimate the ground energy to precision $h$ by running the controlled-time-evolution of the Hamiltonian for $\wt{\Or}(1/(\gamma h^{3/2}))$ \footnote{\label{fn:ge}In \cite{GeTuraCirac2019}, the meaning of the notation $\wt{\Or}(\cdot)$ is different from that in our work. In particular, $\wt{\Or}(\cdot)$ in \cite{GeTuraCirac2019} hides all factors that are poly-logarithmic in $1/h$, $1/\epsilon$, $1/\gamma$, and $1/\Delta$, regardless of what is inside the parentheses. We preserve their notation when citing their results since these factors do not play an important role when comparing the complexities of our methods.}, and querying $U_I$ $\wt{\Or}(1/(\gamma \sqrt{h}))$ times. It is worth noting that their method requires $h=\wt{\Or}(\Delta)$, and therefore is very expensive when the gap is extremely small. When the ground energy is not known \apriori, Ge \etal proposed a method to first estimate the ground energy and then apply the LCU approach.

In recent years several hybrid quantum-classical algorithms have been developed to estimate the ground energy, or to prepare the ground state, or both. The variational quantum eigenvalue solver (VQE) \cite{PeruzzoMcCleanShadboltEtAl2014} has gained much attention recently because of its low requirement for circuit depth and its variational structure. However the exact complexity of this algorithm is not clear because it relies on a proper choice of ansatz and needs to solve a non-convex optimization problem. Other such algorithms include quantum imaginary-time evolution, quantum Lanczos \cite{MottaSunTanEtAl2019}, and quantum filter diagonalization \cite{ParrishMcMahon2019,StairHuangEvangelista2019}. Their complexities are either quasi-polynomial or unknown.


The recent development of block-encoding \cite{BerryChildsKothari2015} and quantum signal processing (QSP) {\cite{LowYoderChuang2016,LowChuang2017,GilyenSuLowEtAl2019}} enables us to apply non-unitary operators, specifically polynomials of a block-encoded matrix efficiently. It uses a minimal number of ancilla qubits, and avoids the Hamiltonian simulation. These will be the basic tools of this work, of which we give a brief introduction below.

Block-encoding is a powerful tool to represent a non-unitary matrix in the quantum circuit. A matrix $A\in\CC^{N\times N}$ where $N=2^n$ can be encoded in the upper-left corner of an $(m+n)$-qubit unitary matrix if 
\begin{equation}
\norm{A-\alpha(\bra{0^m}\otimes I) U (\ket{0^m}\otimes I)}_2\leq \epsilon.
\label{eqn:block_encoding}
\end{equation}
In this case we say $U$ is an $(\alpha,m,\epsilon)$-block-encoding of $A$.
Many matrices of practical interests can be efficiently block-encoded. In particular we will discuss the block-encoding of Hamiltonians of physical systems in Section~\ref{sec:discussions}.

Using the block-encoding of a Hermitian $A$, QSP enables us to construct block-encodings for a large class of polynomial eigenvalue transformations of $A$.  We pay special attention to even or odd polynomials with real coefficients, because we only apply this type of polynomial eigenvalue transformation in this work. Also for simplicity we assume the block-encoding is done without error. {\cite[Theorem 2]{GilyenSuLowEtAl2019} enables us to perform eigenvalue transformation of $A$ for polynomials of definite parity (even or odd)}.
\begin{thm}[QSP for polynomials {of definite parity}]
\label{thm:qsp_parity}
Let $U$ be an $(\alpha,m,0)$-block-encoding of a Hermitian matrix $A$. Let $P\in\RR[x]$ be a degree-$\ell$ even or odd real polynomial and $\abs{P(x)}\le 1$ for any $x\in[-1,1]$. Then there exists an $(1, m+1,0)$-block-encoding $\wt{U}$ of $P(A/\alpha)$ using $\ell$ queries of $U$, $U^{\dagger}$, and $\Or((m+1)\ell)$ other primitive quantum gates. 
\end{thm} 
\begin{rem}
{\cite[Theorem 2]{GilyenSuLowEtAl2019} provides a singular value transformation for any square matrix $A$ and polynomials of definite parity. When $A$ is a Hermitian matrix, the eigenvalue transformation is the same as the singular value transformation \cite[Page 203]{GilyenSuLowEtAl2019}. A related statement in the same paper is \cite[Theorem 31]{GilyenSuLowEtAl2019}, which describes the eigenvalue transformation of a Hermitian matrix for an arbitrary polynomial, by means of a linear combination of two polynomials of even and odd parities respectively.}
\end{rem}

Constructing the quantum circuit for QSP requires computing a sequence of phase factors beforehand, and there are classical algorithms capable of doing this \cite{Haah2019}. Some recent progress has been made to efficiently compute phase factors for high-degree polynomials to high precision \cite{ChaoDingGilyenEtAl2020,DongMengWhaleyLin2020}. In this work, {unless otherwise specified}, we assume the phase factors are computed without error.



Using the tools introduced above, we assume the Hamiltonian $H$ is given in its $(\alpha,m,0)$-block-encoding $U_H$. This, together with $U_I$, are the two oracles we assume we are given in this work.
QSP enables us to filter eigenstates using fewer qubits than LCU. In \cite{LinTong2019} a filtering method named optimal eigenstate filtering is introduced. It is based on an explicitly constructed optimal minimax polynomial, and achieves the same asymptotic complexity, ignoring poly-logarithmic factors, as the method by Ge \etal when applied to the ground state preparation problem if the ground energy is known exactly.



In this work we first develop a filtering method that filters out all eigenstates corresponding to eigenvalues above a certain threshold. This filtering method enables us to prepare the ground state of a Hamiltonian with spectral gap bounded away from zero when only an upper bound of the ground energy is known, unlike in the filtering methods discussed above which all require either exact value or high-precision estimate of the ground energy. Our filtering method has an exponentially improved dependence on precision compared to Kitaev's phase estimation \cite{Kitaev1995} and uses fewer qubits compared to other variants of the phase estimation algorithm \cite{PoulinWocjan2009,GeTuraCirac2019}. This filtering method, applied to the initial state given in our assumption, also enables us to tell whether the ground energy is smaller than $a$ or greater than $b$ for some $b>a$, with high probability. Therefore a binary search yields a ground energy estimate with success probability arbitrarily close to one. We then combine the filtering method and ground energy estimation to prepare the ground state when no non-trivial bound for the ground energy is known. A comparison of the query complexities between the method in our work and the corresponding ones in \cite{GeTuraCirac2019}, which to our best knowledge achieve state-or-the-art query complexities, are shown in Table~\ref{tab:compare_algs}.
 \begin{table}[ht!]
    \centering
    \makegapedcells
        \begin{tabular}{|p{10mm}|c|p{30mm}|p{32mm}|p{32mm}|}
        \hline
        \hline
          &   & Preparation (bound known) & Ground energy & Preparation (bound unknown) \\
        \hline
        \multirow{2}{*}{$U_H$} & This work & $\Or\left(\frac{\alpha}{\gamma\Delta}\log(\frac{1}{\epsilon})\right)$  & $\wt{\Or}\left(\frac{\alpha}{\gamma h}\log(\frac{1}{\vartheta})\right)$  &  $\wt{\Or}\left(\frac{\alpha}{\gamma \Delta}\log(\frac{1}{\vartheta\epsilon})\right)$  \\  
        \cline{2-5}
         & Ge \etal & $\wt{\Or}\left(\frac{\alpha}{\gamma\Delta}\right)$  & $\wt{\Or}\left(\frac{\alpha^{3/2}}{\gamma h^{3/2}}\right)$  &  $\wt{\Or}\left(\frac{\alpha^{3/2}}{\gamma \Delta^{3/2}}\right)$  \\  [10pt]
        \hline
        \multirow{2}{*}{$U_I$} & This work & $\Or\left(\frac{1}{\gamma}\right)$ & $\wt{\Or}\left(\frac{1}{\gamma}\log(\frac{\alpha}{h})\log(\frac{1}{\vartheta})\right)$  &  $\wt{\Or}\left(\frac{1}{\gamma}\log(\frac{\alpha}{\Delta})\log(\frac{1}{\vartheta})\right)$  \\  
        \cline{2-5}
         & Ge \etal & $\wt{\Or}\left(\frac{1}{\gamma}\right)$ & $\wt{\Or}\left(\frac{1}{\gamma}\sqrt{\frac{\alpha}{h}}\right)$  &  $\wt{\Or}\left(\frac{1}{\gamma}\sqrt{\frac{\alpha}{\Delta}}\right)$  \\  
        \hline
         Extra & This work &  $\Or(1)$  & $\Or(\log(\frac{1}{\gamma}))$ &  $\Or(\log(\frac{1}{\gamma}))$  \\  
        \cline{2-5}
         qubits & Ge \etal & $\Or(\log(\frac{1}{\Delta}\log(\frac{1}{\epsilon})))$  & $\Or(\log(\frac{1}{h}))$  & $\Or(\log(\frac{1}{\Delta}\log(\frac{1}{\epsilon})))$    \\  
        \hline
        \hline
        \end{tabular}
    \caption{The query complexities of algorithms and number of extra qubits used in our work and the corresponding ones by Ge \etal in \cite{GeTuraCirac2019}. $\alpha,\gamma,\Delta,\epsilon$ are the same as above and $h$ is the precision of the ground energy estimate. By extra qubits we mean the ancilla qubits that are not part of the block-encoding. In this work the ground energy estimation algorithm and the algorithm to prepare ground state without \apriori bound have success probabilities lower bounded by $1-\vartheta$, while in \cite{GeTuraCirac2019} the corresponding algorithms have constant success probabilities. The complexities for algorithms by Ge \etal are estimated assuming Hamiltonian simulation is done as in \cite{LowChuang2019}. The usage of the notation $\wt{\Or}$ is  \cite{GeTuraCirac2019} different from that in our work, as explained in footnote~\ref{fn:ge}.}
    \label{tab:compare_algs}
\end{table}

From the query complexities in Table~\ref{tab:compare_algs} we can see our method for ground energy estimation achieves a exponential speedup in terms of the dependence of number of queries to $U_I$ on the ground energy estimate precision $h$ and a speedup of $1/\sqrt{h}$ factor in the  dependence of number of queries to $U_H$ on the precision. Moreover, Ge \etal assumes in their work that the precision $h=\wt{\Or}(\Delta)$, while we make no such assumptions. This gives our algorithm even greater advantage when the gap is much smaller than desired precision. This becomes useful in the case of preparing a low energy state (not necessarily a ground state). Because Ge \etal used a slightly different query assumption, \ie access to time-evolution rather than block-encoding, when computing the complexities for methods in \cite{GeTuraCirac2019} in Table~\ref{tab:compare_algs} we assume the Hamiltonian simulation is done with $\Or(\alpha t)$ queries to $U_H$, and the error is negligible. This can be achieved using the Hamiltonian simulation in \cite{LowChuang2019}, and cannot be asymptotically improved because of the complexity lower bound proved in \cite{BerryChildsKothari2015}. Therefore the comparison here is fair even though our work makes use of a different oracle.  Also \cite{GeTuraCirac2019} assumed a scaled Hamiltonian $H$ with its spectrum contained in $[0,1]$. We do not make such an assumption, and therefore the $\alpha$ factor should be properly taken into account as is done in Table~\ref{tab:compare_algs}.


\vspace{1em}
\noindent\textbf{Organization:}
The rest of the paper is organized as follows. In Section~\ref{sec:ref_and_proj} we use QSP to construct block-encodings of reflectors and projectors associated with eigen-subspaces. In Section~\ref{sec:with_bound} we use the projectors to prepare ground state when an upper bound of the ground energy is given. In Section~\ref{sec:wo_bound} we introduce the ground energy estimation algorithm, a hybrid quantum-classical algorithm based on the binary search, and use it to prepare the ground state when no ground energy upper bound is known \apriori. In Section~\ref{sec:optimality} we show the dependence of our query complexities on the overlap and gap is essentially optimal by considering the unstructured search problem. We also show the dependence of our ground energy estimation algorithm on the precision is nearly optimal by considering the quantum approximate counting problem. In Section~\ref{sec:low_energy} we use our methods to prepare low-energy states when the spectral lower gap is unknown, or even when the ground state is degenerate. In Section~\ref{sec:discussions} we discuss practical issues and future research directions.

\section{Block-encoding of reflector and projector}
\label{sec:ref_and_proj}

A key component in our method is a polynomial approximation of the sign function in the domain $[-1,-\delta]\cup[\delta,1]$. The error scaling of the best polynomial approximation has been studied in \cite{EremenkoYuditskii2007}, and an explicit construction of a polynomial with the same error scaling is provided in \cite{LowChuang2017} based on the approximation of the $\mathrm{erf}$ function. We quote  \cite[Lemma 14]{GilyenSuLowEtAl2019} here with some small modification:
\begin{lem}[Polynomial approximation of the sign function]
\label{lem:sign_poly}
For all $0<\delta<1$, $0<\epsilon<1$, there exists an efficiently computable odd polynomial $S(\cdot;\delta,\epsilon)\in\RR[x]$ of degree $\ell=\Or(\frac{1}{\delta}\log(\frac{1}{\epsilon}))$, such that
\begin{itemize}
\item[(1)] for all $x\in [-1,1]$, $|S(x;\delta,\epsilon)|\leq 1$, and
\item[(2)] for all $x\in[-1,-\delta]\cup[\delta,1]$, $|S(x;\delta,\epsilon)-\mathrm{sign}(x)|\leq \epsilon$.
\end{itemize}
\end{lem}
\begin{rem}
{Compared to \cite[Lemma 14]{GilyenSuLowEtAl2019} we have rescaled the interval from $[-2,2]$ to $[-1,1]$, and this does not result in any substantial change.}
\end{rem}

When we have the $(\alpha,m,0)$-block-encoding of a Hermitian matrix $H=\sum_{k}\lambda_k\ket{\psi_k}\bra{\psi_k}\in\CC^{N\times N}$, $N=2^n$, $\lambda_k\leq \lambda_{k+1}$, we can construct a $(\alpha+|\mu|,m+1,0)$-block-encoding of matrix $H-\mu I$ using  of \cite[Lemma 29]{GilyenSuLowEtAl2019} for any $\mu\in\RR$. Then using QSP, by Theorem~\ref{thm:qsp_parity}, we can obtain an $(1,m+2,0)$-block-encoding of $-S(\frac{H-\mu I}{\alpha+|\mu|};\delta,\epsilon)$ for any $\delta$ and $\epsilon$. If we assume further that $\Delta/2 \leq \min_k |\mu-\lambda_k|$, then we let $\delta=\frac{\Delta}{4\alpha}$, and by Lemma~\ref{lem:sign_poly} all the eigenvalues of $-S(\frac{H-\mu I}{\alpha+|\mu|};\delta,\epsilon)$ are $\epsilon$-close to either 0 or 1. Therefore $-S(\frac{H-\mu I}{\alpha+|\mu|};\delta,\epsilon)$ is $\epsilon$-close, in operator norm, to the reflector about the direct sum of eigen-subspaces corresponding to eigenvalues smaller than $\mu$:
\[
R_{<\mu} = \sum_{k:\lambda_k<\mu} \ket{\psi_k}\bra{\psi_k} - \sum_{k:\lambda_k>\mu} \ket{\psi_k}\bra{\psi_k},
\]
and thus the block-encoding is also an $(1,m+2,\epsilon)$-block-encoding of $R_{<\mu}$. We denote this block-encoding by $\REF(\mu,\delta,\epsilon)$. We omitted the dependence on $H$ because $H$ as well as its block-encoding is usually fixed in the rest of the paper.

{In the above discussion we have used QSP in a black-box manner. For concreteness, we present a single-qubit illustrative example to demonstrate how to use a block-encoded Hamiltonian to construct the reflector in Appendix~\ref{sec:numerical_example}.}

Because our goal is to prepare the ground state, we will use the projector more often than the reflector. Now we construct a block-encoding of projector using $\REF(\mu,\delta,\epsilon)$ by the following circuit
\begin{equation}
\label{eq:circ_proj}
\Qcircuit @C=1em @R=.3em{
\lstick{\ket{0}}           & \qw & \gate{\mathrm{H}} & \ctrl{1}   & \gate{\mathrm{H}} & \qw\\
\lstick{\ket{0^{m+2}}} & \qw & \qw      & \multigate{1}{\REF(\mu,\delta,\epsilon)} & \qw & \qw \\
\lstick{\ket{\phi}}    & \qw & \qw      & \ghost{\REF(\mu,\delta,\epsilon)} & \qw & \qw
}
\end{equation}
where $\mathrm{H}$ is the Hadamard gate, and we denote this circuit as $\PROJ(\mu,\delta,\epsilon)$. 
{Note that
\[
\begin{aligned}
&(\bra{0^{m+3}}\otimes I)\PROJ(\mu,\delta,\epsilon)(\ket{0^{m+3}}\otimes I) \\
&= \Big(\bra{+}\bra{0^{m+2}}\otimes I\Big)\Big(\ket{0}\bra{0}\otimes I\otimes I + \ket{1}\bra{1}\otimes \REF(\mu,\delta,\epsilon)\Big)\Big(\ket{+}\ket{0^{m+2}}\otimes I\Big) \\
&= \frac{1}{2}\Big(I+(\bra{0^{m+2}}\otimes I)\REF(\mu,\delta,\epsilon)(\ket{0^{m+2}}\otimes I)\Big),
\end{aligned}
\]
and we have
\[
\begin{aligned}
&\|(\bra{0^{m+3}}\otimes I)\PROJ(\mu,\delta,\epsilon)(\ket{0^{m+3}}\otimes I)-P_{<\mu}\| \\
&\leq \frac{1}{2}\|(\bra{0^{m+2}}\otimes I)\REF(\mu,\delta,\epsilon)(\ket{0^{m+2}}\otimes I)-R_{<\mu}\| \\
&\leq \frac{\epsilon}{2}.
\end{aligned}
\]
}Here $P_{<\mu}$ is the projector into the direct sum of eigen-subspaces corresponding to eigenvalues smaller than $\mu$
\[
P_{<\mu} = \sum_{k:\lambda_k<\mu} \ket{\psi_k}\bra{\psi_k}
{=\frac{1}{2}(I+R_{<\mu})}.
\]
Therefore $\PROJ(\mu,\delta,\epsilon)$ is an $(1,m+3,\epsilon/2)$-block-encoding of $P_{<\mu}$. In fact this can still be seen as an application of linear combination of block encoding \cite[Lemma~29]{GilyenSuLowEtAl2019}, using the relation $P_{<\mu}=\frac{1}{2}(R_{<\mu}+I)$.

We use the following lemma to summarize the results
\begin{lem}[Reflector and projector]
\label{lem:be_ref_proj}
Given a Hermitian matrix $H$ with its $(\alpha,m,0)$-block-encoding $U_H$, with the guarantee that $\mu\in\RR$ is separated from the spectrum of $H$ by a gap of at least $\Delta/2$, we can construct an $(1,m+2,\epsilon)$-block-encoding of $R_{<\mu}$, and an $(1,m+3,\epsilon/2)$-block-encoding of $P_{<\mu}$, both using $\Or(\frac{\alpha}{\Delta}\log(\frac{1}{\epsilon}))$ applications of $U_H$ and $U_H^\dagger$, and $\Or(\frac{m\alpha}{\Delta}\log(\frac{1}{\epsilon}))$ other one- and two-qubit gates.
\end{lem}

We remark that for the block-encoding $\PROJ(\mu,\delta,\epsilon)$, even a failed application of it can give us potentially useful information. We have
\[
\PROJ(\mu,\delta,\epsilon)\ket{0^{m+3}}\ket{\phi}=\ket{0}\ket{0^{m+2}}P_{<\mu}\ket{\phi}+\ket{1}\ket{0^{m+2}}P_{>\mu}\ket{\phi}+\frac{1}{\sqrt{2}}\ket{{-}}\ket{E},
\]
where $P_{>\mu} = I- P_{<\mu}$ and $\ket{E}$ satisfies $\|\ket{E}\|\leq \epsilon$. Thus when we apply the block-encoding and measure the first two registers, \ie the first $m+3$ qubits, we have probability at least $1-\frac{\epsilon^2}{2}$ to obtain an outcome with either 0 or 1 followed by $(m+2)$ 0's. In the former case the projection has been successful, and in the latter case we have obtained an approximation of $P_{>\mu}\ket{\phi}$. 

If we do not treat the output of 1 followed by $m+2$ 0's as failure then there is another interpretation of the circuit $\PROJ(\mu,\delta,\epsilon)$: this is an approximate projective measurement $\{P_{<\mu},P_{>\mu}\}$. In fact the whole circuit can be seen as phase estimation on a reflector, which needs only one ancilla qubit. 

\section{Algorithm with \apriori ground energy bound}
\label{sec:with_bound}

With the approximate projector developed in the previous section we can readily design an algorithm to prepare the ground state. We assume we have the Hamiltonian $H$ given through its block-encoding as in the last section. If we are further given an initial state $\ket{\phi_0}$ prepared by a unitary $U_{I}$, \ie $U_{I}\ket{0^n}=\ket{\phi_0}$, and the promises that for some known $\gamma>0$, $\mu$, and $\Delta$, we have
\begin{enumerate}[label=\textnormal{(P\arabic*)}]
\item \label{overlap_bound} Lower bound for the overlap: $|\braket{\phi_0|\psi_0}|\geq \gamma$,
\item \label{gap_bound} Bounds for the ground energy and spectral gap: $\lambda_0 \leq \mu - \Delta/2 < \mu + \Delta/2 \leq \lambda_1$.
\end{enumerate}
Here $\mu$ is an upper bound for the ground energy, $\Delta$ is a lower bound for the spectral gap, and $\gamma$ is a lower bound for the initial overlap. Now suppose we want to prepare the ground state to precision $\epsilon$, we can use Lemma~\ref{lem:be_ref_proj} to build a block-encoding of the projector $P_{<\mu}=\ket{\psi_0}\bra{\psi_0}$, and then apply it to $\ket{\phi_0}$ which we can prepare. This will give us something close to $\ket{\psi_0}$. We use fidelity to measure how close we can get. To achieve $1-\epsilon$ fidelity we need to use circuit $\PROJ(\mu,\Delta/4\alpha,\gamma\epsilon)$, and we denote,
\[
\wt{P}_{<\mu} = (\bra{0^{m+3}}\otimes I)\PROJ(\mu,\Delta/4\alpha,\gamma\epsilon)(\ket{0^{m+3}}\otimes I)
\]
then the resulting fidelity will be
\[
\frac{|\braket{\psi_0|\wt{P}_{<\mu}|\phi_0}|}{\|\wt{P}_{<\mu}\ket{\phi_0}\|} \geq \frac{|\braket{\psi_0|\phi_0}|-\gamma\epsilon/2}{|\braket{\psi_0|\phi_0}|+\gamma\epsilon/2} \geq
1-\frac{\gamma\epsilon}{|\braket{\psi_0|\phi_0}|}\geq 1-\epsilon.
\]
Here we have used
\[
\|\wt{P}_{<\mu}\ket{\phi_0}\|\le \norm{P_{<\mu}\ket{\phi_0}+(\wt{P}_{<\mu}-P_{<\mu})\ket{\phi_0}}\le |\braket{\psi_0|\phi_0}|+\gamma\epsilon/2.
\]
This is when we have a successful application of the block-encoding. The success probability is
\[
\|\wt{P}_{<\mu}\ket{\phi_0}\|^2 \geq \left(\|P_{<\mu}\ket{\phi_0}\|-\frac{\gamma\epsilon}{2}\right)^2 \geq \gamma^2\left(1-\frac{\epsilon}{2}\right)^2.
\]
With amplitude amplification \cite{BrassardHoyerMoscaEtAl2002} we can boost the success probability to $\Omega(1)$ with $\Or(\frac{1}{\gamma})$ applications of $\PROJ(\mu,\Delta/4\alpha,\gamma\epsilon)$ and its inverse, as well as $\Or(\frac{m}{\gamma})$ other one- and two- qubit gates. Here we are describing the expected complexity since the procedure succeeds with some constant probability. In amplitude amplification we need to use a reflector similar to the oracle used in Grover's search algorithm \cite{Grover1996fast}. Instead of constructing a reflector from $\PROJ(\mu,\Delta/4\alpha,\gamma\epsilon)$ we can directly use $\REF(\mu,\Delta/4\alpha,\gamma\epsilon)$ constructed in the previous section.

We summarize the results in the following theorem
\begin{thm}[Ground state preparation with \apriori ground energy bound]
\label{thm:with_bound}
Suppose we have Hamiltonian $H=\sum_{k}\lambda_k \ket{\psi_k}\bra{\psi_k}\in\CC^{N\times N}$, where $\lambda_k\leq \lambda_{k+1}$, given through its $(\alpha,m,0)$-block-encoding $U_H$. Also suppose we have an initial state $\ket{\phi_0}$ prepared by circuit $U_{I}$, as well as the promises \ref{overlap_bound} and \ref{gap_bound}. Then the ground state $\ket{\psi_0}$ can be prepared to fidelity $1-\epsilon$ with the following costs:
\begin{enumerate}
\item Query complexity: $\Or(\frac{\alpha}{\gamma\Delta}\log(\frac{1}{\gamma\epsilon}))$ queries to $U_H$ and $\Or(\frac{1}{\gamma})$ queries to $U_{I}$,
\item Number of qubits: $\Or(n+m)$,
\item Other one- and two- qubit gates: $\Or(\frac{m\alpha}{\gamma\Delta}\log(\frac{1}{\gamma\epsilon}))$.
\end{enumerate}
\end{thm}

\section{Algorithm without \apriori ground energy bound}
\label{sec:wo_bound}

Next we consider the case when we are not given a known $\mu$ to bound the ground energy from above.  All other assumptions about $H$ and its eigenvalues and eigenstates are identical to the previous sections. The basic idea is to test different values for $\mu$ and perform a binary search. This leads to a quantum-classical hybrid method that can estimate the ground energy as well as preparing the ground state to high precision.

All eigenvalues must be in the interval $[-\alpha,\alpha]$, thus we first partition $[-\alpha,\alpha]$ by grid points $-\alpha=x_0<x_1<\ldots<x_{G}=\alpha$, where $x_{k+1}-x_k=h$ for all $k$. 
 Then we attempt to locate $\lambda_0$ in a small interval between two grid points (not necessarily adjacent, but close) through a binary search. To do a binary search we need to be able to tell whether a given $x_k$ is located to the left or right of $\lambda_0$. Because of the random nature of measurement we can only do so correctly with some probability, and we want to make this probability as close to 1 as possible. This is achieved using a technique we call binary amplitude estimation.

\begin{lem}[Binary amplitude estimation]
\label{lem:bae}
Let $U$ be a unitary that acts on two registers, the first register indicating success or failure. Let $A=\|(\bra{0}\otimes I)U(\ket{0}\ket{0})\|$ be the success amplitude. Given $\gamma_0$ and $\gamma_1$, $\Delta := \gamma_1-\gamma_0>0$, provided that $A$ is either smaller than $\gamma_0$ or greater than $\gamma_1$, we can correctly distinguish between the two cases, i.e. output $0$ for the former and $1$ for the latter, with probability $1-\delta$ using $\mathcal{O}((1/\Delta)\log(1/\delta))$ applications of (controlled-) $U$ and its inverse.
\end{lem}
\begin{proof}
The proof is essentially identical to the proof for gapped phase estimation in \cite{Ambainis2012, ChildsKothariSomma2017}. We can perform amplitude estimation up to error $\Delta/4$ with $\mathcal{O}(1/\Delta)$ applications of $U$ and $U^\dagger$. This has a success probability of $8/\pi^2$ according to Theorem 12 of \cite{BrassardHoyerMoscaEtAl2002}. We turn the estimation result into a boolean indicating whether it is larger or smaller than $(\gamma_0+\gamma_1)/2$. The boolean is correct with probability at least $8/\pi^2$. Then we do a majority voting to boost this probability. Chernoff bound guarantees that to obtain a $1-\delta$ probability of getting the correct output we need to repeat $\mathcal{O}(\log(1/\delta))$ times. Therefore in total we need to run $U$ and $U^\dagger$ $\mathcal{O}((1/\Delta)\log(1/\delta))$ times.
\end{proof}

We then apply binary amplitude estimation to the block-encoding of the projector defined in \eqref{eq:circ_proj} $\PROJ(x_k,h/2\alpha,\epsilon')$ for some precision $\epsilon'$ to be chosen.
We denote the amplitude of the ``good'' component after applying block-encoding by
$$
A_k=\|(\bra{0^{m+3}}\otimes I)\PROJ(x_k,h/2\alpha,\epsilon')(\ket{0^{m+3}}\ket{\phi})\|,
$$
which satisfies the following:
\[
A_{k}\begin{cases}
\geq \gamma - \frac{\epsilon'}{2}, & \lambda_{0}\leq x_{k-1},\\
\leq \frac{\epsilon'}{2}, & \lambda_{0}\geq x_{k+1}.
\end{cases}
\]
We can then let 
\[
\epsilon'=\gamma/2,
\]
the two amplitudes are separated by a gap lower bounded by $\gamma/2$. Therefore we can run the binary amplitude estimation, letting $U$ in Lemma~\ref{lem:bae} be 
$$
U=\PROJ(x_k,h/2\alpha,\epsilon')(I\otimes U_I),
$$
to correctly distinguish the two cases where $\lambda_0\leq x_{k-1}$ and $\lambda_0 \geq x_{k+1}$ with probability $1-\delta$, by running $\PROJ(x_k,h/2\alpha,\epsilon')$, $U_I$, and their inverses $\mathcal{O}((1/\gamma)\log(1/\delta))$ times. The output of the binary amplitude estimation is denoted by $B_k$.

We then define $\mathcal{E}$ as the event that an error occurs in the final result of binary amplitude estimation when we are computing $B_k$ for some $k$ such that $x_{k+1}<\lambda_0$ or $x_{k-1}>\lambda_0$ in our search process. All future discussion is conditional on $\mathcal{E}^c$ meaning that there is no error in binary amplitude estimation for $B_k$ when $x_{k+1}<\lambda_0$ or $x_{k-1}>\lambda_0$. This has a probability that is at least $(1-\delta)^R$ where $R$ is the number of times binary amplitude estimation is run.

Conditional on $\mathcal{E}^c$, almost surely (with probability 1) $B_k=1$ when $\lambda_0\leq x_{k-1}$ and $B_k=0$ when $\lambda_0\geq x_{k+1}$. Therefore $B_k=0$ tells us $\lambda_0>x_{k-1}$ and $B_k=1$ tells us $\lambda_0<x_{k+1}$. $B_k$ and $B_{k+1}$ combined give us the information as shown in Table~\ref{tab:lambda0_pos}.
\begin{table}[h!]
\centering
\begin{tabular}{|c|c|c|}
\hline 
\hline 
$B_k$ & $B_{k+1}$ & Position of $\lambda_0$ \\ 
\hline 
1 & 1 & $\lambda_0<x_{k+1}$ \\ 
\hline 
0 & 0 & $\lambda_0>x_k$ \\ 
\hline 
0 & 1 & $x_{k-1}<\lambda_0<x_{k+2}$ \\ 
\hline 
1 & 0 & $x_{k}<\lambda_0<x_{k+1}$ \\ 
\hline 
\hline 
\end{tabular} 
\caption{Conditional on $\mathcal{E}^c$, $B_k$ and $B_{k+1}$ can provide us with the information as shown in the table.}
\label{tab:lambda0_pos}
\end{table}

\begin{algorithm}[ht!]
\caption{Binary search to locate $\lambda_0$}
\label{alg:binary_search}
\begin{algorithmic}
\STATE $L \gets 0$, $U\gets G$
\WHILE{$U-L>3$}
\STATE $k=\lfloor (L+U)/2 \rfloor$
\STATE Run binary amplitude estimation to get $B_k$ and $B_{k+1}$.
\SWITCH{$B_k,B_{k+1}$}
\CASELINE{$(1,1)$} $U\gets k+1$
\CASELINE{$(0,0)$} $L\gets k$
\CASELINE{$(0,1)$} \textbf{return} $k-1$, $k+2$
\CASELINE{$(1,0)$} \textbf{return} $k$, $k+1$
\ENDSWITCH
\ENDWHILE
\RETURN $L$, $U$
\end{algorithmic}
\end{algorithm}

Using the Table~\ref{tab:lambda0_pos} we can do the binary search as outlined in Algorithm~\ref{alg:binary_search}. 
{For the $\ell$-th step in Algorithm~\ref{alg:binary_search} we denote the integer variables $U$ and $L$ by $U_\ell$ and $L_\ell$. In all four outcomes for $(B_k,B_{k+1})$, if the algorithm does not terminate at this step, then the new $U_{\ell+1}-L_{\ell+1}$ will be at most $(U_\ell-L_\ell)/2+1$. Since $U_0-L_0=G$ at the very beginning, we can show inductively $U_\ell-L_\ell\leq (G-2)/2^\ell+2$. Therefore when $\ell \geq \log_2(G-2)$ we have $U_\ell-L_\ell\leq 3$. Thus}
the algorithm must terminate in $\lceil \log_2(G) \rceil=\mathcal{O}(\log(\alpha/h))$ steps. The output we denote by $L$ and $U$. They satisfy $x_L<\lambda_0<x_U$ and $U-L\leq 3$.

If we want the whole procedure to be successful with probability at least $1-\vartheta$, then we need $\mathrm{Prob}(\mathcal{E}^c)\geq 1-\vartheta$. Since 
\[
\mathrm{Prob}(\mathcal{E}^c)\geq (1-\delta)^{\lceil \log_2(G) \rceil}\geq (1-\delta)^{\log_2(4\alpha/h)},
\]
we only need, for small $\vartheta$,
\[
\delta \leq \frac{\vartheta}{2\log_2(4\alpha/h)}.
\]

Algorithm~\ref{alg:binary_search} enables us to locate $\lambda_0$ within an interval of length at most $3h$. In total we need to run binary amplitude estimation at most $\mathcal{O}(\log(\alpha/h))$ times. Each amplitude estimation queries $\PROJ(x_k,h/2\alpha,\epsilon')$ and $U_I$ $\Or((1/\gamma)\log(1/\delta))$ times, where $\epsilon'=\gamma/2$. Therefore the number of queries to $U_H$ and $U_I$ are respectively
\[
\Or\left(\frac{\alpha}{\gamma h}\log\left(\frac{\alpha}{h}\right)\log\left(\frac{1}{\gamma}\right)\log\left(\frac{\log(\alpha/h)}{\vartheta}\right)\right), \quad
\Or\left(\frac{1}{\gamma}\log\left(\frac{\alpha}{h}\right)\log\left(\frac{\log(\alpha/h)}{\vartheta}\right)\right).
\]

In particular, in the procedure above we did not use \ref{gap_bound} but only used \ref{overlap_bound}. Therefore we do not need to assume the presence of a gap. The result can be summarized into the following theorem:
\begin{thm}[Ground energy]
\label{thm:ground_energy}
Suppose we have Hamiltonian $H=\sum_{k}\lambda_k \ket{\psi_k}\bra{\psi_k}\in\CC^{N\times N}$, where $\lambda_k\leq \lambda_{k+1}$, given through its $(\alpha,m,0)$-block-encoding $U_H$. Also suppose we have an initial state $\ket{\phi_0}$ prepared by circuit $U_{I}$, as well as the promise \ref{overlap_bound}. Then the ground energy can be estimated to precision $h$ with probability $1-\vartheta$ with the following costs:
\begin{enumerate}
\item Query complexity: $\Or\left(\frac{\alpha}{\gamma h}\log\left(\frac{\alpha}{h}\right)\log\left(\frac{1}{\gamma}\right)\log\left(\frac{\log(\alpha/h)}{\vartheta}\right)\right)$ queries to $U_H$ and \\ $\Or\left(\frac{1}{\gamma}\log\left(\frac{\alpha}{h}\right)\log\left(\frac{\log(\alpha/h)}{\vartheta}\right)\right)$ queries to $U_{I}$,
\item Number of qubits: $\Or(n+m+\log(\frac{1}{\gamma}))$,
\item Other one- and two- qubit gates: $\Or\left(\frac{m\alpha}{\gamma h}\log\left(\frac{\alpha}{h}\right)\log\left(\frac{1}{\gamma}\right)\log\left(\frac{\log(\alpha/h)}{\vartheta}\right)\right)$.
\end{enumerate} 
\end{thm}

The extra $\Or(\log(1/\gamma))$ qubits needed come from amplitude estimation, which uses phase estimation. If we use Kitaev's original version of phase estimation using only a single qubit \cite{Kitaev1995}, we can reduce the number of extra qubits to $\Or(1)$.
With Theorem~\ref{thm:ground_energy} we can then use Algorithm~\ref{alg:binary_search} to prepare the ground state without knowing an upper bound for the ground energy beforehand, when in addition to \ref{overlap_bound} we have a lower bound for the spectral gap:
\begin{enumerate}[label=\textnormal{(P2')}]
\item \label{gap_bound_weak} Bound for the spectral gap: $\lambda_1-\lambda_0\geq \Delta$.
\end{enumerate}
We first run Algorithm~\ref{alg:binary_search} to locate the ground energy in an interval $[x_L,x_U]$ of length at most $\Delta$. Then we simply apply $\PROJ((x_L+x_U)/2,\Delta/4\alpha,\gamma\epsilon)$ to $\ket{\phi_0}$. This will give us an approximate ground state with at least $1-\epsilon$ fidelity. Therefore we have the following corollary:
\begin{cor}[Ground state preparation without \apriori bound]
\label{cor:without_bound}
Suppose we have Hamiltonian  $H=\sum_{k}\lambda_k \ket{\psi_k}\bra{\psi_k}\in\CC^{N\times N}$, where $\lambda_k\leq \lambda_{k+1}$, given through its $(\alpha,m,0)$-block-encoding $U_H$. Also suppose we have an initial state $\ket{\phi_0}$ prepared by circuit $U_{I}$, as well as the promises \ref{overlap_bound} and \ref{gap_bound_weak}. Then the ground state can be can be prepared to fidelity $1-\epsilon$ with probability $1-\vartheta$ with the following costs:
\begin{enumerate}
\item Query complexity: $\Or\left(\frac{\alpha}{\gamma \Delta}\left(\log\left(\frac{\alpha}{\Delta}\right)\log\left(\frac{1}{\gamma}\right)\log\left(\frac{\log(\alpha/\Delta)}{\vartheta}\right)+\log\left(\frac{1}{\epsilon}\right)\right)\right)$ queries to $U_H$ and  $\Or\left(\frac{1}{\gamma}\log\left(\frac{\alpha}{\Delta}\right)\log\left(\frac{\log(\alpha/\Delta)}{\vartheta}\right)\right)$ queries to $U_{I}$,
\item Number of qubits: $\Or(n+m+\log(\frac{1}{\gamma}))$,
\item Other one- and two- qubit gates: $\Or\left(\frac{m\alpha}{\gamma \Delta}\left(\log\left(\frac{\alpha}{\Delta}\right)\log\left(\frac{1}{\gamma}\right)\log\left(\frac{\log(\alpha/\Delta)}{\vartheta}\right)+\log\left(\frac{1}{\epsilon}\right)\right)\right)$.
\end{enumerate} 
\end{cor}

It may be sometimes desirable to ignore whether the procedure is successful or not. In this case we will see the output as a mixed state whose density matrix is
\[
\rho = \mathrm{Prob}(\mathcal{E}^c)\ket{\wt{\psi}_0}\bra{\wt{\psi}_0} + \rho',
\]
where $\ket{\wt{\psi}_0}$ is the approximate ground state with fidelity at least $1-\epsilon$, which is produced conditional on the event $\mathcal{E}^c$, and $\Tr\rho'=\mathrm{Prob}(\mathcal{E})$. Then this mixed state will have a fidelity lower bounded by
\[
\braket{\psi_0|\rho|\psi_0} \geq  \mathrm{Prob}(\mathcal{E}^c) |\braket{\wt{\psi}_0|\psi_0}|^2\geq (1-\vartheta)(1-\epsilon)^2.
\]
If we want to achieve $\sqrt{1-\xi}$ fidelity for the mixed state, we can simply let $\vartheta=\epsilon=\xi/3$. Thus the number of queries to $U_H$ and $U_I$ are $\wt{\Or}(\frac{\alpha}{\gamma\Delta}\log(\frac{1}{\xi}))$ and $\wt{\Or}(\frac{1}{\gamma}\log(\frac{\alpha}{\Delta})\log(\frac{1}{\xi}))$ respectively.

\section{Optimality of the query complexities}
\label{sec:optimality}

In this section we prove for the ground state preparation algorithms outlined in Section~\ref{sec:with_bound} and Section~\ref{sec:wo_bound} the number of queries to $U_H$ and $U_I$ are essentially optimal. We will also show our ground energy estimation algorithm has an nearly optimal dependence on the precision. We first prove the following complexity lower bounds:
\begin{thm}
\label{thm:optimal}
Suppose we have a generic Hamiltonian $H=\sum_{k}\lambda_k \ket{\psi_k}\bra{\psi_k}\in\CC^{N\times N}$, where $\lambda_k\leq \lambda_{k+1}$, given through its $(\alpha,m,0)$-block-encoding $U_H$, and $\alpha=\Theta(1)$. Also suppose we have an initial state $\ket{\phi_0}$ prepared by circuit $U_{I}$, as well as the promises \ref{overlap_bound} and \ref{gap_bound}. Then the query complexities of preparing the ground state $\ket{\psi_0}$ of $H$ to fidelity at least $\sqrt{3}/2$ satisfy
\begin{enumerate}
\item When $\Delta=\Omega(1)$, and $\gamma\rightarrow 0^+$, the number of queries to $U_H$ is $\Omega(1/\gamma)$;
\item When $\gamma=\Omega(1)$, and $\Delta\rightarrow 0^+$, the number of queries to $U_H$ is $\Omega(1/\Delta)$;
\item When $\Delta=\Omega(1)$, and $\gamma\rightarrow 0^+$, {it is not possible to accomplish the above task using $\Or(1/\gamma^{1-\theta})$ queries to $U_I$ and $\Or(\mathrm{poly}(1/\gamma))$ queries to $U_H$ for any $\theta>0$.}
\end{enumerate}
\end{thm}

\begin{proof}
We prove all three lower bounds by applying the ground state preparation algorithm to the unstructured search problem. In the unstructured search problem we try to find a $n$-bit string $t$ marked out by the oracle
\[
U_t = I - 2\ket{t}\bra{t}.
\]
It is proved for this problem the number of queries to $U_t$ to find $t$ with probability $1/2$ is lower bounded by $\Omega(\sqrt{N})$ where $N=2^n$  \cite{BennettBernsteinBrassardVazirani1997}.

This problem can be seen as a ground state preparation problem. We find that $\ket{t}$ is the ground state of $U_t$, which is at the same time a unitary and therefore an $(1,0,0)$-block-encoding of itself. Therefore $U_t$ serves as the $U_H$ in the theorem. The spectral gap is $2$. Also, let
\[
\ket{u}=\frac{1}{\sqrt{N}}\sum_s\ket{s}
\]
be the uniform superposition of all $n$-strings, then we have
$
\braket{u|t}=\frac{1}{\sqrt{N}}
$,
and $\ket{u}$ can be efficiently prepared by the Hadamard transform since $\mathrm{H}^{\otimes n}\ket{0^n}=\ket{u}$. Therefore $\mathrm{H}^{\otimes n}$ serves as the $U_I$ described in the theorem.

If the ground state preparation problem can be solved with $o(1/\gamma)$ queries to $U_H$ for fixed $\Delta$ to produce an approximate ground state with fidelity at least $\sqrt{3}/2$, then from the above setup we have $\gamma=1/\sqrt{N}$, and we can first find the approximate ground state and then measure in the computational basis, obtaining $t$ with probability at least $3/4$. Therefore the unstructured search problem can be solved with $o(\sqrt{N})$ queries to the oracle $U_t$, which is impossible. Thus we have proved the first lower bound in our theorem.

To prove the second lower bound we want to create a situation in which the overlap is bounded from below by a constant but the gap vanishes. We need to introduce the Grover diffusion operator
\begin{equation}
\label{eq:grover_diffusion}
D = I_n-2\ket{u}\bra{u}.
\end{equation}
which can be efficiently implemented. Then we define
\begin{equation}
\label{eq:hamilton_comb}
H(\tau) = (1-\tau)D + \tau U_t,
\end{equation}
and consider $H(1/2)$. 
{Because both $\operatorname{span}(\ket{u},\ket{t})$ and its orthogonal complement are invariant subspaces of $D$ and $U_t$, and both operators become the identity operator when restricted to the orthogonal complement of $\operatorname{span}(\ket{u},\ket{t})$, we only need to look for the ground state in the 2-dimensional subspace $\operatorname{span}(\ket{u},\ket{t})$. In this subspace, relative to the basis $\{\ket{u},\ket{t}\}$, the matrix representation of $H(1/2)$ is
\[
\begin{pmatrix}
0 & -\braket{u|t} \\
-\braket{t|u} & 0
\end{pmatrix}
=-\frac{1}{\sqrt{N}}\begin{pmatrix}
0 & 1 \\
1 & 0
\end{pmatrix}.
\]
}Therefore the ground state of $H(1/2)$ is 
\[
\ket{\Psi}=\frac{\ket{u}+\ket{t}}{\sqrt{2+\frac{2}{\sqrt{N}}}}.
\]
and therefore $\braket{\Psi|u}=\braket{\Psi|t}= 1/\sqrt{2}+\Or(1/\sqrt{N})$ for large $N$. Furthermore, the gap is $\Delta(1/2)=2/\sqrt{N}$.

Therefore $\ket{t}$ can be prepared in the following way: we first prepare the ground state of $H(1/2)$, whose block-encoding is easy to construct using one application of $U_t$. The resulting approximate ground state we denote by $\ket{\wt{\Psi}}$. Then we measure $\ket{\wt{\Psi}}$ in the computational basis. If there is some non-vanishing probability of obtaining $t$ then we can boost the success probability to above $1/2$ by repeating the procedure and verifying using $U_t$.

If the second lower bound in the theorem does not hold, then $\ket{\wt{\Psi}}$ can be prepared with $o(1/\Delta(1/2))=o(\sqrt{N})$ queries to the block-encoding of $H(1/2)$ and therefore the same number of queries to $U_t$. Because the angle corresponding to fidelity is the great-circle distance on the unit sphere, we have the triangle inequality
(using that $|\braket{\wt{\Psi}|\Psi}|\ge \sqrt{3}/{2}$) 
\[
\arccos |\braket{\wt{\Psi}|t}| \leq \arccos |\braket{\Psi|t}| + \arccos |\braket{\wt{\Psi}|\Psi}|\le \frac{5\pi}{12} + \Or\left(\frac{1}{\sqrt{N}}\right).
\]
Therefore for large $N$ we have $|\braket{\wt{\Psi}|t}|\geq \cos(5\pi/12) + \Or(1/\sqrt{N})>1/4$. The probability of getting $t$ when performing measurement is at least $1/16$. Therefore we can boost the success probability to above $1/2$ by $\Or(1)$ repetitions and verifications. The total number of queries to $U_t$ is therefore $o(\sqrt{N})$. Again, this is impossible. Therefore we have proved the second lower bound in our theorem.

For the last lower bound we need to create some trade off between the gap and the overlap. We consider preparing the ground state of the Hamiltonian $H(1/2-N^{-1/2+\delta})$, $0<\delta<1/6$, whose block-encoding can be efficiently constructed with a single application of $U_t$, as an intermediate step. It is shown in Appendix~\ref{app:search} that the ground state is
\begin{equation}
\label{eq:interm_state_lb3}
\ket{\Phi}=\ket{u} + \frac{1}{4}N^{-\delta}\ket{t} + \Or(N^{-2\delta}).
\end{equation}
Therefore
\[
\gamma_u=|\braket{\Phi|u}| = 1+\Or(N^{-2\delta}),\quad \gamma_t=|\braket{\Phi|t}| = \frac{1}{4}N^{-\delta} + \Or(N^{-2\delta}).
\]
Also we show in Appendix~\ref{app:search} that the gap is
\begin{equation}
\label{eq:gap_lb3}
\Delta(1/2-N^{-1/2+\delta})=4N^{\delta-1/2}+\Or(N^{-1/2-\delta}).
\end{equation}

We first apply the algorithm described in Section~\ref{sec:with_bound} to prepare the ground state of $H(1/2-N^{-1/2+\delta})$ to fidelity $1-N^{-2\delta}/128$. Using the overlap $\gamma_u$ and the gap in \eqref{eq:gap_lb3}, the approximate ground state, denoted by $\ket{\wt{\Phi}}$, can be prepared with $\Or(N^{1/2-\delta}\log(N))$ queries to the block-encoding of $H(1/2-N^{-1/2+\delta})$, and therefore the same number of queries to $U_t$.

The overlap between $\ket{\wt{\Phi}}$ and $\ket{t}$ can again be bounded using the triangle inequality
\[
\begin{aligned}
\arccos|\braket{\wt{\Phi}|t}| &\leq \arccos|\braket{\Phi|t}| + \arccos|\braket{\wt{\Phi}|\Phi}| \\
&\leq \arccos\left(\frac{N^{-\delta}}{4}\right) + \arccos\left(1-\frac{N^{-2\delta}}{128}\right) + \Or(N^{-2\delta}) \\
&\leq \frac{\pi}{2} - \frac{N^{-\delta}}{4} + \sqrt{2\times \frac{N^{-2\delta}}{128}} + \Or(N^{-2\delta}) \\
&=\frac{\pi}{2} - \frac{N^{-\delta}}{8} + \Or(N^{-2\delta}).
\end{aligned}
\]
Therefore we have
\[
\wt{\gamma}_t = |\braket{\wt{\Phi}|t}| \geq \frac{N^{-\delta}}{8} + \Or(N^{-2\delta}).
\]

If the last lower bound in our theorem does not hold, we can then prepare the ground state of $U_t$ by using the initial state $\ket{\wt{\Phi}}$ only $\Or(1/\wt{\gamma}^{1-\theta}_t)$ times for some $\theta>0$, and the number of queries to $U_t$ at this step, \ie not including the queries used for preparing $\ket{\wt{\Phi}}$, is $\Or(1/\wt{\gamma}^p_t)$ for some $p>0$. Therefore the total number of queries to $U_t$ is
\[
\Or\left(\frac{N^{1/2-\delta}\log(N)}{\wt{\gamma}^{1-\theta}_t}+\frac{1}{\wt{\gamma}^p_t}\right)=\Or(N^{1/2-\delta\theta}\log(N)+N^{\delta p}).
\]
This complexity must be $\Omega(N^{1/2})$ according to the lower bound for unstructured search problem. Therefore we need $\delta p\geq 1/2$. However we can choose $\delta$ to be arbitrarily small, and no finite $p$ can satisfy this condition. Hence we have a contradiction. This proves the last lower bound in our theorem.
\end{proof}

When we look at the query complexities of the ground state preparation algorithms in Secs.~\ref{sec:with_bound} and \ref{sec:wo_bound}, we can use $\wt{\Or}$ notation to hide the logarithmic factors, and both algorithms use $\wt{\Or}(\frac{\alpha}{\gamma\Delta})$ queries to $U_H$ and $\wt{\Or}(\frac{1}{\gamma})$ queries to $U_I$ when we want to achieve some fixed fidelity. Given the lower bound in Theorem~\ref{thm:optimal} we can see the algorithm with \apriori bound for ground energy essentially achieves the optimal dependence on $\gamma$ and $\Delta$. The algorithm without \apriori bound for ground energy achieves the same complexity modulo logarithmic factors, while using less information. This fact guarantees that the dependence is also nearly optimal.

We will then prove the nearly optimal dependence of our ground energy estimation algorithm on the precision $h$. We have the following theorem:
\begin{thm}
\label{thm:optimal_energy}
Suppose we have a generic Hamiltonian $H=\sum_{k}\lambda_k \ket{\psi_k}\bra{\psi_k}\in\CC^{N\times N}$, where $\lambda_k\leq \lambda_{k+1}$, given through its $(\alpha,m,0)$-block-encoding $U_H$, and $\alpha=\Theta(1)$. Also suppose we have an initial state $\ket{\phi_0}$ prepared by circuit $U_{I}$, as well as the promise that $|\braket{\phi_0|\psi_0}|=\Omega(1)$. Then estimating the ground energy to precision $h$ requires $\Omega(1/h)$ queries to $U_H$.
\end{thm}

This time we convert the quantum approximate counting problem, which is closely related to the unstructured search problem, into an eigenvalue problem. The quantum approximate counting problem is defined in the following way. We are given a set of $n$-bit strings $S\subset\{0,1\}^n$ specified by the oracle $U_f$ satisfying
\[
U_f\ket{x} = \begin{cases}
-\ket{x} & \ x \in S, \\
\ket{x} & \ x \notin S,
\end{cases}
\]
for any $x\in\{0,1\}^n$. We want to estimate the size $|S|/N$ up to relative error $\epsilon$. It has been proven that this requires $\Omega\left(\frac{1}{\epsilon}\sqrt{\frac{N}{|S|}}\right)$ queries to $U_f$ for $|S|=o(N)$ \cite[Theorem 1.13]{NayakWu1999}, where $N=2^n$, for the success probability to be greater than $3/4$, and this lower bound can be achieved using amplitude estimation \cite{BrassardHoyerMoscaEtAl2002}.

We convert this problem into an eigenvalue problem of a block-encoded Hamiltonian. Let $\ket{u}$ be the uniform superposition of the computational basis and $D$ be the Grover diffusion operator defined in \eqref{eq:grover_diffusion}.
Then define the following $(n+1)$-qubit unitary ($\mathrm{H}$ is the Hadamard gate)
\[
U_H = (\mathrm{H}\otimes I_n)[\ket{0}\bra{0}\otimes D-\ket{1}\bra{1}\otimes (U_fDU_f)](\mathrm{H}\otimes I_n),
\]
which can be implemented using two applications of controlled-$U_f$. We define
\[
H = (\bra{0}\otimes I_n)U_H(\ket{0}\otimes I_n)=\frac{1}{2}(D-U_f D U_f).
\]
Note that here $H$ is given in its $(1,1,0)$-block-encoding $U_H$.
Let
\[
\ket{u}=a\ket{u_0} + \sqrt{1-a^2}\ket{u_1}
\]
where the unit vectors $\ket{u_0}$ and $\ket{u_1}$ satisfy
\[
U_f\ket{u_0}=-\ket{u_0},\quad U_f\ket{u_1}=\ket{u_1},
\]
then we find $a=\sqrt{|S|/N}$. We only need to estimate the value of $a$ to precision $\Or(\epsilon'\sqrt{N/|S|})$ in order to estimate $|S|/N$ to precision $\epsilon'$.

We analyze the eigenvalues and eigenvectors of $H$. It can be verified that $\{\ket{u_0},\ket{u_1}\}$ span an invariant subspace of $H$, and relative to this orthonormal basis $H$ is represented by the matrix
\[
\begin{pmatrix}
0 & -2a\sqrt{1-a^2} \\
-2a\sqrt{1-a^2} & 0
\end{pmatrix}.
\]
In the orthogonal complement of this subspace, $H$ is simply the zero matrix. Therefore $H$ has only two non-zero eigenvalues $\pm 2a\sqrt{1-a^2}$ corresponding to eigenvectors
\[
\ket{\psi_{\mp}}=\frac{1}{\sqrt{2}}(\ket{u_0}\mp\ket{u_1}).
\]
The ground state of $H$ is therefore $\ket{\psi_+}$ with ground energy $-2a\sqrt{1-a^2}$. We can use $\ket{u}$ as the initial state, with an overlap $\braket{\psi_+|u}=\frac{1}{\sqrt{2}}(a+\sqrt{1-a^2})\geq\frac{1}{\sqrt{2}}$.

We use this Hamiltonian to prove Theorem \ref{thm:optimal_energy}:
\begin{proof}
Assume toward contradiction that there exists an algorithm that estimates the ground energy to precision $h$ using only $o(1/h)$ queries to $U_H$. Then we use this algorithm to estimate the ground energy of the block-encoded Hamiltonian constructed above, for $a=o(1)$, which means $|S|=o(N)$. Estimating $2a\sqrt{1-a^2}$ to precision $\Or(h)$ enables us to estimate $a$ to precision $\Or(h)$. Setting $h=\epsilon'\sqrt{N/|S|}$, then this algorithm can estimate $|S|/N$ to precision $\epsilon'$, with success probability at least $3/4$. Since we are interested in the relative error we set $\epsilon'=\epsilon|S|/N$. Therefore the whole procedure uses only $o(1/h)=o(\frac{1}{\epsilon}\sqrt{\frac{N}{|S|}})$ queries to $U_H$ and therefore twice the amount of queries to $U_f$. This contradicts the lower bound for the approximate counting problem in \cite{NayakWu1999}.
\end{proof}

\begin{rem}
{
Theorem~\ref{thm:optimal_energy} can also be viewed as a consequence of the optimality of the quantum phase estimation algorithm \cite{bessen2005lower}. If instead of the block-encoding $U_H$ we have $e^{-i\tau H}$ as the oracle for some $\tau$ such that $|\tau|\|H\|\leq \pi$, then even when given the exact ground state of $H$, \cite[Lemma 3]{bessen2005lower} gives a query complexity lower bound $\Omega(1/h)$ for estimating the ground energy to within additive error $h$. This provides a different proof of the above theorem, since $e^{-i\tau H}$ and the block-encoding of $H$ are interconvertible: one can efficiently implement $e^{-i\tau H}$ via Hamiltonian simulation starting from a block-encoding of $H$ \cite{LowChuang2019}, and can efficiently obtain a block-encoding of $H$ by querying $e^{-i\tau H}$ according to \cite[Corollary 71]{GilyenSuLowWiebe2018Long}.
}
\end{rem}

\section{Low-energy state preparation}
\label{sec:low_energy}

It is known that estimating the spectral gap $\Delta$ is a difficult task \cite{Ambainis2014physical, CubittPerezGarciaEtAl2015,BauschCubittEtAl2018}. Our algorithm for finding ground energy, as discussed in Theorem~\ref{thm:ground_energy}, does not depend on knowing the spectral gap. However both of our algorithms for preparing the ground state in Theorem~\ref{thm:with_bound} and Corollary~\ref{cor:without_bound} require a lower bound of the spectral gap. We would like to point out that if we only want to produce a low-energy state $\ket{\psi}$, making $\braket{\psi|H|\psi}\leq \mu$ for some $\mu > \lambda_0$, as in \cite{PoulinWocjan2009}, then this can be done without any knowledge of the spectral gap. In fact this is even possible for when the ground state is degenerate.

To do this, we need to first assume we have a normalized initial state $\ket{\phi_0}$ with non-trivial overlap with the low-energy eigen-subspaces. Quantitatively this means for some $\gamma,\delta>0$, if we expand the initial state in the eigenbasis of $H$, obtaining $\ket{\phi_0}=\sum_k \alpha_k \ket{\psi_k}$, then
$
\sum_{k:\lambda_k \leq \mu-3\delta}|\alpha_k|^2\geq \gamma^2.
$
Then we can use the block-encoded projection operator in \eqref{eq:circ_proj} to get
\[
\ket{\psi'} = (\bra{0^{m+3}}\otimes I)\PROJ(\mu-2\delta,\delta,\epsilon')(\ket{0^{m+3}}\otimes \ket{\phi_0}),
\]
for some precision $\epsilon'$. Now we expand $\ket{\psi'}$ in the eigenbasis to get $\ket{\psi'}=\sum_{k}\beta_k\ket{\psi_k}$, and denote $\ket{\varphi'}=\sum_{k:\lambda_k<\mu-\delta}\beta_k\ket{\psi_k}$. We then have, because of the approximation to the sign function,
\[
\|\ket{\psi'}-\ket{\varphi'}\|\leq\frac{\epsilon'}{2},\quad \braket{\varphi'|\varphi'}\geq \gamma^2(1-\frac{\epsilon'}{2})^2,\quad \braket{\varphi'|H|\varphi'}\leq (\mu-\delta)\braket{\varphi'|\varphi'}.
\]
From the above bounds we further get
\[
\frac{\braket{\psi'|H|\psi'}}{\braket{\psi'|\psi'}} \leq \frac{\braket{\varphi'|H|\varphi'}+\|H\|\epsilon'+\|H\|\epsilon'^2/4}{\braket{\varphi'|\varphi'}-\epsilon'} \leq \frac{\mu - \delta + \frac{\alpha\epsilon'+\alpha\epsilon'^2/4}{\gamma^2(1-\epsilon'/2)^2}}{1-\frac{\epsilon'}{\gamma^2(1-\epsilon'/2)^2}}.
\]
Now denoting $\ket{\psi}=\ket{\psi'}/\|\ket{\psi'}\|$ we can make $\braket{\psi|H|\psi}\leq \mu$ by choosing $\epsilon'=\Or(\gamma^2\delta/\alpha)$. Therefore the total number of queries to $U_H$ required is $\Or(\frac{1}{\delta\gamma}\log(\frac{\alpha}{\delta\gamma}))$ and the number of queries to $U_I$ is $\Or(\frac{1}{\gamma})$.

From this we can see that if the initial state $\ket{\phi_0}$ has a overlap with the the ground state that is at least $\gamma$, and we want to prepare a state with energy upper bounded by $\lambda_0+\delta$, the required number of queries to $U_H$ and $U_I$ are $\Or(\frac{1}{\delta\gamma}\log(\frac{\alpha}{\delta\gamma}))$ and $\Or(\frac{1}{\gamma})$ respectively. If we do not know the ground energy beforehand we can use the algorithm in Theorem~\ref{thm:ground_energy} to estimate it first. Note that none of these procedures assumes a spectral gap.


\section{Discussions}
\label{sec:discussions}

In this work we proposed an algorithm to prepare the ground state of a given Hamiltonian when a ground {energy} upper bound is known (Theorem~\ref{thm:with_bound}), an algorithm to estimate the ground energy based on binary search (Theorem~\ref{thm:ground_energy}), and combining these two to get an algorithm to prepare the ground state without knowing an upper bound \apriori (Corollary~\ref{cor:without_bound}). By solving the unstructured search problem and the approximate counting problem through preparing the ground state, we proved that the query complexities for the tasks above cannot be substantially {improved}, as otherwise the complexity lower bound for the two problems would be violated.

All our algorithms are based on the availability of the block-encoding of the target Hamiltonian. This is a non-trivial task but we know it can be done for many important settings. For example, Childs \etal proposed an LCU approach to block-encode the Hamiltonian of a quantum spin system \cite{ChildsMaslovEtAl2018}, in which the Hamiltonian is decomposed into a sum of Pauli matrices. In \cite{LowWiebe2018}, Low and Wiebe outlined the methods to construct block-encoding of Hubbard Hamiltonian with long-range interaction, and of quantum chemistry Hamiltonian in plane-wave basis, both using fast-fermionic Fourier transform (FFFT) \cite{BabbushWiebeEtAl2018}. The FFFT can be replaced by a series of Givens rotations which gives lower circuit depth and better utilizes limited connectivity \cite{KivlichanMcCleanEtAl2018,JiangSungEtAl2018}. {Any sparse Hamiltonian whose entries can be efficiently computed can also be block-encoded using a quantum walk operator \cite{BerryChilds2009black,BerryChildsKothari2015,ChildsKothariSomma2017}. }

We remark that the quantum circuit used in our method for ground energy estimation can be further simplified. The main obstacle to applying this method to near-term devices is the need of amplitude estimation, which requires phase estimation. It is possible to replace amplitude estimation by estimating the success probability classically. In the context of binary amplitude estimation in Lemma~\ref{lem:bae}, we need to determine whether the success amplitude is greater than $3\gamma/4$ or smaller than $\gamma/4$. This can be turned into a classical hypothesis testing to determine whether the success probability is greater than $9\gamma^2/16$ or smaller than $\gamma^2/16$. A simple Chernoff bound argument tells us that we  need $\Or(\log(1/\vartheta)/\gamma^2)$ samples to distinguish the two cases with success probability at least $1-\vartheta$, as opposed to the $\Or(\log(1/\vartheta)/\gamma)$ complexity in amplitude estimation.

In this approach, the only quantum circuit we need to use is the one in \eqref{eq:circ_proj}. The circuit depth is therefore only $\Or((\alpha/h)\log(1/\gamma))$. It also does not require the $\Or(\log(1/\gamma))$ qubits that are introduced as a result of using amplitude estimation. These features make it suitable for near-to-intermediate term devices.

In \cite{LinTong2019} we proposed an eigenstate filtering method (similar in spirit to the method proposed in Section~\ref{sec:with_bound}),  and we combined it with quantum Zeno effect \cite{ChildsDeottoFarhiEtAl2002, BoixoKnillSomma2009} to solve the quantum linear system problem. The resulting algorithm utilizes the fact that the desired eigenstate along the eigenpath always corresponds to the eigenvalue 0. In the setting of quantum Zeno effect based state preparation, in which we have a series of Hamiltonians and wish to incrementally prepare the ground state of each of them, our algorithm in Theorem~\ref{thm:with_bound} can be used to go from the ground state of one Hamiltonian to the next one, provided that we have a known upper bound for the ground energy. In the absence of such an upper bound, there is the possibility of using the algorithm in Corollary~\ref{cor:without_bound} to solve this problem. However in this setting we only want to use the initial state once for every Hamiltonian, since preparing the initial state involves going through the ground state of all previous Hamiltonians. This presents a challenge and is a topic for our future work.

It is worth pointing out that none of the Hamiltonians used in the proofs of lower bounds in Section~\ref{sec:optimality} is a local Hamiltonian, and therefore our lower bounds do not rule out the possibility that if special properties such as locality are properly taken into consideration, better complexities can be achieved.

\section*{Acknowledgements}
 This work was partially supported by the Department of Energy under Grant No. DE-SC0017867, the Quantum Algorithm Teams Program under Grant No. DE-AC02-05CH11231, the Google Quantum Research Award (L.L.), and by the Air Force Office of Scientific Research under award number FA9550-18-1-0095 (L.L. and Y.T.). We thank Andr{\'a}s Gily{\'e}n, {and the anonymous referees} for helpful suggestions.
 
\bibliographystyle{abbrvnat}
\bibliography{ground_state}

\appendix

\section{{An example of block-encoding and constructing the reflector}}
\label{sec:numerical_example}

{
In this section we use $\sigma_x$, $\sigma_y$, and $\sigma_z$ to denote the three Pauli matrices. We use $\mathrm{H}$ to denote the Hadamard gate. We consider a single-qubit illustrative example of block-encoded matrix and obtain the corresponding reflector through QSP. 
}

{
The matrix we consider is 
\[
H(a) = a \sigma_x + (1-a) I,
\]
for $0\leq a\leq 1$.  Its block-encoding can be using the following circuit
\[
 \Qcircuit @C=0.8em @R=1.em {
 & \gate{V(a)} & \ctrlo{1}  & \gate{V^\dagger(a)} & \qw \\
 & \qw & \gate{\sigma_x}  & \qw & \qw \\
}
\]
where
\[
V(a)=
\begin{pmatrix}
\sqrt{a} & -\sqrt{1-a} \\
\sqrt{1-a} & \sqrt{a}
\end{pmatrix}.
\]
We denote the above circuit by $U_H(a)$. This is a $(\alpha,m,0)$-block-encoding of $H(a)$ where $\alpha=1$ and $m=1$, since we can readily check that
\[
(\bra{0}\otimes I)U_H(a)(\ket{0}\otimes I) = H(a).
\]
}

{
The eigendecomposition of $H(a)$ is
\[
H(a) = \ket{+}\bra{+} + (1-2a)\ket{-}\bra{-},
\]
with eigenvalues $\lambda_{+}(a) = 1,\lambda_{-}(a) = 1-2a$. Our goal is to implement the reflector
\[
R_{<0}(a) = -\sign(H(a)) = -\ket{+}\bra{+} - \sign(1-2a)\ket{-}\bra{-}.
\]
To do this we need an odd polynomial $S(x;\delta,\epsilon)$ introduced in Lemma~\ref{lem:sign_poly}. Instead of the construction done in Ref.~\cite{LowChuang2017} we use the Remez algorithm \cite{remes1934calcul} to obtain this polynomial. We choose $\delta=0.2$ and the $L^{\infty}$ error of the residual is required to be less than $10^{-4}$, i.e. $\epsilon \leq 10^{-4}$.
}

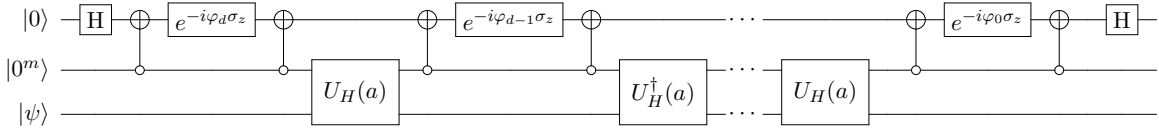
\begin{figure}[htb]
\begin{center}
{
\[
\scalebox{0.8}{
\Qcircuit @C=0.8em @R=1.em {
 \lstick{\ket{0}}& \gate{\mathrm{H}} & \targ & \gate{e^{-i \varphi_{d} \sigma_z}} & \targ & \qw & \targ & \gate{e^{-i \varphi_{d-1} \sigma_z}} & \targ & \qw & \qw &\raisebox{0em}{$\cdots$}&&  \qw & \qw   &\targ & \gate{e^{-i \varphi_0 \sigma_z}} & \targ & \qw  & \gate{\mathrm{H}}&\qw  \\
\lstick{\ket{0^m}}& \qw  &\ctrlo{-1} & \qw  & \ctrlo{-1} & \multigate{1}{U_H(a)} & \ctrlo{-1} & \qw & \ctrlo{-1} & \multigate{1}{U^{\dag}_H(a)} &\qw &\raisebox{0em}{$\cdots$} && \multigate{1}{U_H(a)} & \qw    &\ctrlo{-1} & \qw & \ctrlo{-1} & \qw& \qw & \qw \\
\lstick{\ket{\psi}}& \qw &\qw &\qw&\qw &\ghost{U_H(a)} &\qw&\qw&\qw&\ghost{U_H(a)}&\qw &\raisebox{0em}{$\cdots$} && \ghost{U_H(a)} & \qw&\qw&\qw   & \qw&\qw& \qw &\qw  
}
}
\]
}
\end{center}
\caption{{The circuit implementing the polynomial eigenvalue transformation through QSP for an odd polynomial with phase factors $\{\varphi_j\}_{j=0}^d$. $\mathrm{H}$ is the Hadamard gate and $\sigma_z$ is the Pauli-$Z$ gate.}}
\label{fig:qsp_circuit_real}
\end{figure}

{
Given the polynomial $S(x;\delta,\epsilon)$, using the optimization method proposed in Ref.~\cite{DongMengWhaleyLin2020}, we find a polynomial $P(x)\in \CC [x]$ of odd degree $d$ such that 
\[
\max_{x\in[-1,1]}|\Re P(x)-S(x;\delta,\epsilon)|\leq \epsilon',
\]
where $P(x)$ is characterized by a sequence of phase factors $\{\varphi_j\}_{j=0}^d$ satisfying
\begin{equation}
\label{eq:phase_fac_P}
\begin{pmatrix}
P(x) & \cdot \\
\cdot & \cdot
\end{pmatrix}
=
e^{i\varphi_0 \sigma_z}\prod_{j=1}^d[R(x)e^{i\varphi_j \sigma_z}],
\end{equation}
where 
\[
R(x)=
\begin{pmatrix}
x & \sqrt{1-x^2} \\
\sqrt{1-x^2} & -x
\end{pmatrix}.
\]
The existence of the phase factors is guaranteed by \cite[Theorem 5]{GilyenSuLowEtAl2019}. Ref.~\cite{DongMengWhaleyLin2020} uses quasi-Newton method to solve a least squares problem to obtain these phase factors, and we terminate the iteration only when  $L^\infty$ error of the residual of the real part is smaller than $\epsilon'=10^{-4}$.
}

{
The circuit in Figure~\ref{fig:qsp_circuit_real} with phase factors $\{\varphi_j\}_{j=0}^d$ implements the transformation $H/\alpha\mapsto \Re P(H/\alpha)\approx S(H/\alpha;\delta,\epsilon)$. The various components of this circuit are explained in detail in \cite[Figure 3]{GilyenSuLowEtAl2019}. An important component of this circuit is
\[
\Qcircuit @C=0.8em @R=1.em {
 & \targ & \gate{e^{-i \varphi \sigma_z}} & \targ & \qw \\
 & \ctrlo{-1} & \qw & \ctrlo{-1} & \qw \\
}
\]
where the first register has one qubit, the second register has $m$-qubits, and the open bullet indicates control-on-zero for multiple control qubits. This component implements the operator
\[
 \ket{0}\bra{0}\otimes (e^{i \varphi (2\ket{0^m}\bra{0^m}-I)}) + \ket{1}\bra{1}\otimes (e^{-i \varphi (2\ket{0^m}\bra{0^m}-I)}).
\]
For a detailed discussion see \cite[Corollary 11]{GilyenSuLowEtAl2019}.
}
 

\begin{figure}[ht]
    \centering
    \includegraphics[width=0.5\textwidth]{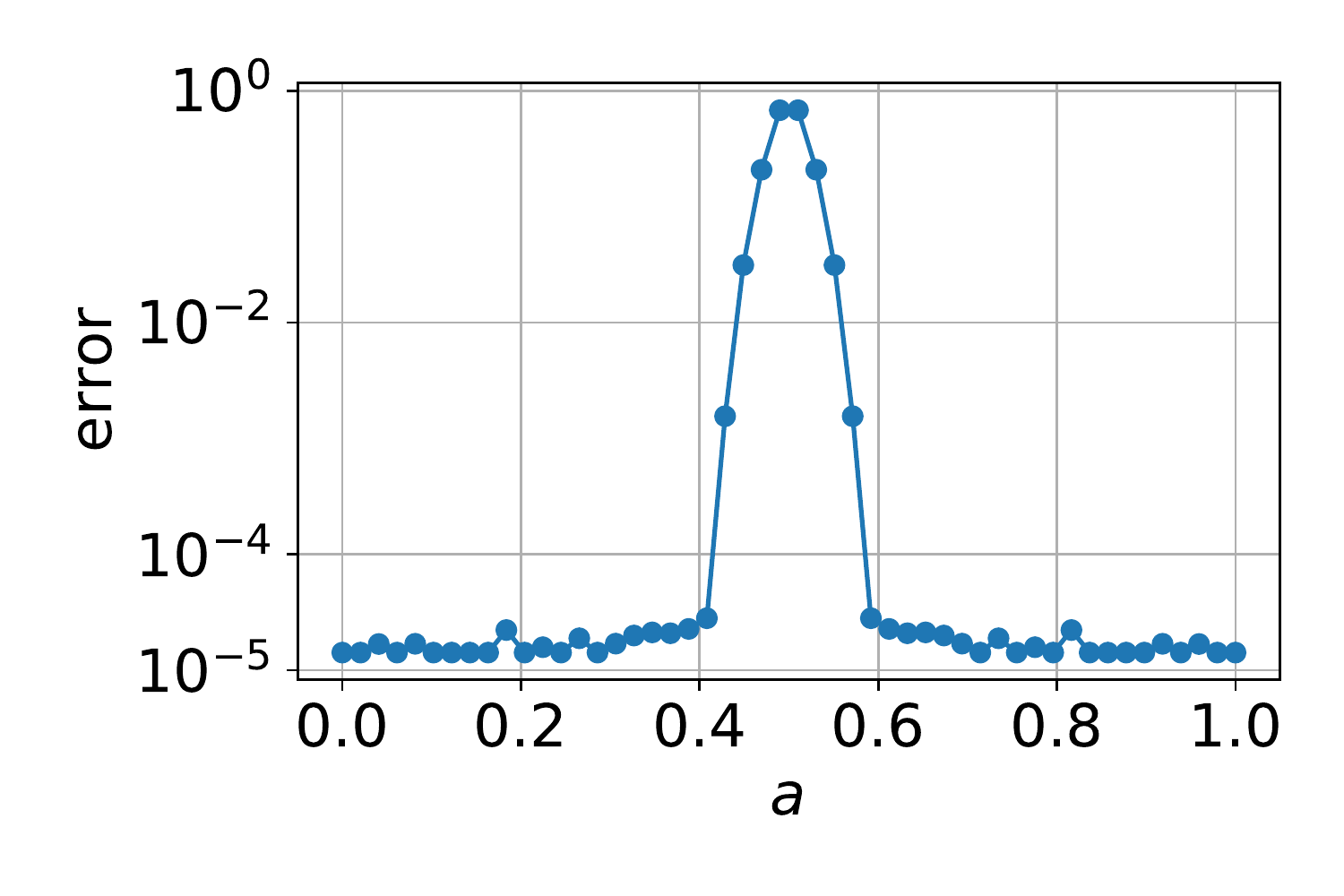}
    \caption{{The error of implementing $R_{<0}(a)$ for $a\in[0,1]$ using QSP, with polynomial $S(x;\delta,\epsilon)$ where $\delta=0.2$ and $\epsilon$ is of the order of $10^{-4}$. The vertical axis uses logarithmic scale. }}
    \label{fig:err_plot}
\end{figure}

{
Using the above circuit, Lemma~\ref{lem:be_ref_proj} guarantees that when the eigenvalues of $H(a)$ are contained in $[-1,-\delta]\cup[\delta,1]$, we will have a good approximation of $R_{<0}(a)$. However when at least one eigenvalue, which in our case can only be $\lambda_-(a)=1-2a$, is in $(-\delta,\delta)$, or in other words when $a\in (0.4,0.6)$,  there is no such guarantee. We plot the operator norm error between the approximate reflector obtained through QSP and the exact reflector $R_{<0}(a)$ in Figure~\ref{fig:err_plot}. It can be seen in the figure that the error is smaller than $10^{-4}$ everywhere except for $a\in(0.4,0.6)$, where the error spikes.
}

\section{Gap and overlap in the unstructured search problem}
\label{app:search}
In this appendix we compute the spectral gap of the Hamiltonian $H(1/2-N^{-1/2+\delta})$ for $H(\tau)$ defined in \eqref{eq:hamilton_comb}, $0<\delta<1/6$, and the overlap between its ground state and $\ket{u}$ and $\ket{t}$ defined in Section~\ref{sec:optimality}.

The first thing we should realize is that we only need to care about the subspace of the Hilbert space spanned by $\ket{u}$ and $\ket{t}$. In the orthogonal complement of this subspace $H(\tau)$ is  simple a multiple of identity. In this subspace, with respect to the non-orthogonal basis $\{\ket{u},\ket{t}\}$, the operator $H(1/2-N^{-1/2+\delta})$ is represented by the following matrix
\begin{equation}
\label{eq:ham_search}
N^{\delta-1/2}
\begin{pmatrix}
-2 & -(N^{-\delta}+2N^{-1/2}) \\
-(N^{-\delta}-2N^{-1/2}) & 2
\end{pmatrix}.
\end{equation}
Direct calculation shows the eigenvalues 
are
\[
\lambda_{\pm} = \pm N^{\delta-1/2}\sqrt{4+N^{-2\delta}-4N^{-1}} = \pm N^{\delta-1/2}(2+\frac{1}{4}N^{-2\delta}+\Or(N^{-4\delta})).
\]
Thus we obtain the spectral gap in \eqref{eq:gap_lb3}. To simplify notation we let $\wt{\lambda}=N^{1/2-\delta}\lambda_+$. We then compute the ground state. We first find an eigenvector corresponding to $\lambda_-$
\[
\begin{aligned}
\ket{\chi} &= N^\delta((N^{-\delta}+2N^{-1/2})\ket{u}+(-2+\wt{\lambda})\ket{t}) \\
&=(1+2N^{\delta-1/2})\ket{u} + (\frac{1}{4}N^{-\delta}+\Or(N^{-3\delta}))\ket{t}  \\
&=\ket{u}+ \frac{1}{4}N^{-\delta}\ket{t}+\Or(N^{\delta-1/2}).
\end{aligned}
\]
We still need to normalize $\ket{\chi}$. The normalization factor is
\[
\begin{aligned}
\|\ket{\chi}\|& = \sqrt{(1+2N^{\delta-1/2})^2 + (\frac{1}{4}N^{-\delta}+\Or(N^{-3\delta}))^2 + \frac{2}{\sqrt{N}}(1+2N^{\delta-1/2})(\frac{1}{4}N^{-\delta}+\Or(N^{-3\delta}))} \\
 &= 1 + \Or(N^{-2\delta}).
\end{aligned}
\]
Note that the third term under the square root comes from the overlap between $\ket{u}$ and $\ket{t}$, and it does not play an important role asymptotically. Therefore normalizing we have the expression for the normalized eigenstate \eqref{eq:interm_state_lb3}.

\end{document}